\documentclass[letter]{aa}

\usepackage{graphicx}
\usepackage{txfonts}
\usepackage{lipsum}
\usepackage{xcolor}
\usepackage{subcaption}
\usepackage{hyperref}
\usepackage{amsmath}
\usepackage{tabularx, multirow}
\usepackage{soul}
\usepackage{hyperref}
\usepackage{booktabs}
\hypersetup{
    colorlinks,
    linkcolor={red!50!black},
    citecolor={blue!50!black},
    urlcolor={blue!80!black}
}

\def\ergs{erg s$^{-1}$}

\def\msun{\ifmmode M_{\odot} \else M$_{\odot}$\fi}
\def\msunyr{\ifmmode M_{\odot} {\rm yr}^{-1} \else M$_{\odot}$ yr$^{-1}$\fi}
\def\msunyrvol{\ifmmode \msunyr {\rm Mpc}^{-3} \else \msunyr Mpc$^{-3}$\fi}
\def\zsun{\ifmmode Z_{\odot} \else Z$_{\odot}$\fi}
\def\lsun{\ifmmode L_{\odot} \else L$_{\odot}$\fi}

\newcommand{\lya}{\ifmmode{\rm Ly}\alpha\else Ly$\alpha$\fi}
\newcommand{\oh}{\ifmmode 12 + \log({\rm O/H}) \else$12 + \log({\rm
O/H})$\fi}

\newcommand{\oiii}{\ifmmode \text{[O~{\sc iii}]} \else[O~{\sc iii}]\fi}

\newcommand{\ha}{\ifmmode \text{H}\alpha \else H$\alpha$\fi}
\newcommand{\hb}{\ifmmode \text{H}\beta \else H$\beta$\fi}
\newcommand{\oiiihb}{\text{\oiii+\hb}}

\newcommand{\oiiia}{\ifmmode\oiii\lambda4363\else$\oiii\lambda4363$\fi}
\newcommand{\OIII}{\ifmmode \oiii\lambda\lambda\lambda4363,4960,5008\else$\oiii\lambda\lambda\lambda4363,4960,5008$\fi}

\def\Oiii{\ifmmode \oiii\lambda\lambda4960,5008 \else\oiii$\lambda\lambda4960,5008$\fi}
\def\Oiiib{\ifmmode \oiii\lambda5008 \else\oiii$\lambda5008$\fi}
\def\Oiiit{\ifmmode \oiii\lambda4960 \else\oiii$\lambda4960$\fi}

\newcommand{\mstar}{\ifmmode M_\star \else $M_\star$\fi}
\newcommand{\muv}{\ifmmode M_\text{UV}\else $M_\text{UV}$\fi}
\newcommand{\auv}{\ifmmode A_{\rm UV} \else $A_{\rm UV}$\fi}
\newcommand{\luv}{\ifmmode L_{\rm UV} \else $L_{\rm UV}$\fi}
\newcommand{\lir}{\ifmmode L_{\rm IR} \else $L_{\rm IR}$\fi}
\newcommand{\lbol}{\ifmmode L_{\rm bol} \else $L_{\rm bol}$\fi}
\newcommand{\liruv}{\ifmmode L_{\rm IR+UV} \else $L_{\rm IR+UV}$\fi}
\newcommand{\liroveruv}{\ifmmode L_{\rm IR}/L_{\rm UV} \else $L_{\rm IR}/L_{\rm UV}$\fi}
\newcommand{\nlyc}{\ifmmode N_{\rm Lyc} \else $N_{\rm Lyc} $\fi}
\newcommand{\rholyc}{\ifmmode \rho_{\rm Lyc} \else $\rho_{\rm Lyc} $\fi}
\newcommand{\chion}{\ifmmode \xi_{\rm ion} \else $\xi_{\rm ion}$\fi}
\newcommand{\chioncorr}{\ifmmode \xi_{\rm ion}^0 \else $\xi_{\rm ion}^0$\fi}
\newcommand{\Rthree}{\ifmmode {\rm R3}\else R3\fi}
\newcommand{\Rthreefunc}{\ifmmode \Rthree(\muv) \else $\Rthree(\muv)$ \fi}

\newcommand{\fesc}{\ifmmode f_\textrm{esc} \else $f_\textrm{esc}$ \fi}
\newcommand{\Z}{\ifmmode 12+\log({\rm O/H}) \else $12+\log({\rm O/H})$ \fi}
\newcommand{\Rt}{\Rthree}
\newcommand{\nion}{\ifmmode\dot{N}_{\rm ion}\else$\dot{N}_{\rm ion}$\fi}
\newcommand{\xion}{\ifmmode\xi_{\rm ion}\else$\xi_{\rm ion}$\fi}

\begin{document} 

    \title{GLIMPSE-D: Metallicity decline in faint galaxies: Implications for \oiiihb\ luminosity function and re-ionisation budget}

    \titlerunning{GLIMPSE-D: Metallicity decline in faint galaxies: \oiiihb\ LF and reionisation budget}

   \author{Damien Korber\inst{1}\thanks{Corresponding author: \href{mailto:damien.korber@protonmail.ch}{damien.korber@protonmail.ch}}
          \and Daniel Schaerer\inst{1,2}
          \and Rui Marques-Chaves\inst{1}
          \and Angela Adamo\inst{14}
          \and Arghyadeep Basu\inst{5}
          \and John Chisholm\inst{9,10}
          \and Miroslava Dessauges-Zavadsky\inst{1}
          \and Kristen.~B.~W. McQuinn\inst{3,4}
          \and Alberto Saldana-Lopez\inst{6}
          \and Hakim Atek\inst{11}
          \and Ryan Endsley\inst{9}
          \and Seiji Fujimoto\inst{7,8}
          \and Lukas J.\ Furtak\inst{9,10}
          \and Vasily Kokorev\inst{9}
          \and Rohan P. Naidu\inst{13}
          \and Richard Pan\inst{12}
          }
   \institute{Department of Astronomy, University of Geneva,
   Chemin Pegasi 51, 1290 Versoix, Switzerland
        \and CNRS, IRAP, 14 Avenue E. Belin, 31400 Toulouse, France
        \and Space Telescope Science Institute, 3700 San Martin Drive, Baltimore, MD, 21218, USA
        \and Rutgers University, Department of Physics and Astronomy, 136 Frelinghuysen Road, Piscataway, NJ 08854, USA
        \and Univ Lyon, Univ Lyon1, Ens de Lyon, CNRS, CRAL UMR5574, F-69230, Saint-Genis-Laval, France
        \and Department of Astronomy, Oskar Klein Centre, Stockholm University, 106 91 Stockholm, Sweden
        \and David A. Dunlap Department of Astronomy and Astrophysics, University of Toronto, Toronto, ON M5S 3H4, Canada
        \and Dunlap Institute for Astronomy and Astrophysics, University of Toronto, Toronto, ON M5S 3H4, Canada
        \and Department of Astronomy, The University of Texas at Austin, Austin, TX 78712, USA
        \and Cosmic Frontier Center, The University of Texas at Austin, Austin, TX 78712, USA
        \and Institut d'Astrophysique de Paris, UMR 7095, CNRS, Sorbonne Université, 98 bis boulevard Arago, 75014 Paris, France
        \and Department of Physics \& Astronomy, Tufts University, MA 02155, USA
        \and MIT Kavli Institute for Astrophysics and Space Research, 70 Vassar Street, Cambridge, MA 02139, USA
        \and Department of Astronomy, The Oskar Klein Centre, Stockholm University, AlbaNova, SE-10691 Stockholm, Sweden
        }

   \date{Received 18 January 2026 / Accepted 11 May 2026}

    \abstract{We report the measurement of the R3=\Oiiib/\hb\ ratios for 54 galaxies in the GLIMPSE-D survey. 
    Thanks to gravitational lensing, our sample includes galaxies with $-20\lesssim\muv<-14$ at $z=6-9$.
    We derived oxygen abundances using calibrated relationships.
    We observe a significant decline in R3 values below $\muv\gtrsim-18$, which we interpret as evidence of decreasing metallicities in fainter regimes.
    We explored four prescription models of the evolution of R3 with UV emission based on the new measurements and results from previous surveys.
    Applying these models to the GLIMPSE \oiiihb\ luminosity functions, we measured and extrapolated the ionising photon-production rate, \nion\, of galaxies down to very faint limits (${\rm SFR}_{\ha}\gtrsim 5\times 10^{-3} \msunyr$).  
    If our results can be generalised, they indicate a dominant contribution of star-forming galaxies to re-ionisation and are consistent with the recent discovery of ultra-faint metal-poor galaxies.
    Our measurements of the relative contribution of each luminosity bin show that galaxies with $L_{\ha}\approx10^{41-42}$\ergs\ dominate at $8<z<9$, but the relative contributions become more uniform at $7<z<8$.
    Extreme models either under- or over-estimate the ionising photon budget, while intermediate models align with recent observational constraints.
    }

   \maketitle\nolinenumbers

\section{Introduction}
During the epoch of re-ionisation, the Universe underwent significant evolution, which shaped its current aspect. This epoch ended around $z\sim5-6$ \citep[e.g.][]{robertson_galaxy_2022} with a fully ionised Universe. 
While main drivers of re-ionisation are well accepted, their relative importance remains uncertain. Some scenarios emphasise the dominant role of faint galaxies \citep[e.g.][]{atek_extreme_2018, simmonds_ionizing_2024}, while others highlight brighter galaxies \citep[e.g.][]{naidu_rapid_2020} or attribute a significant contribution to active galaxy nuclei \citep[AGNs, e.g.][]{singha_faint_2025, grazian_what_2024}.
Constraining the number density of galaxies and ionising photon production 
is essential for understanding their role in cosmic re-ionisation. While previous works have mostly used the UV luminosity functions (LFs) at different redshifts \citep[e.g.][]{bouwens_z_2022}, the relation between the UV luminosity and ionising photon emission is complex. On the other hand, the LF of emission lines provides a more direct measure of the ionising photon production, which can be achieved in the epoch of re-ionisation, for example~with the \oiiihb\ LF at $z \sim 6-9$.

Using the ultra-deep GLIMPSE photometry \citep{atek_jwst_2025}, \cite{korber_glimpse_2026} determined the very faint-end of the \oiiihb~LF, finding an LF flatter than the UV LF, i.e.~effectively a reduced number density of \oiiihb~faint galaxies and a lower ionising photon production than inferred with standard assumptions from the UV LF. To do so, they had to assume a dependence of R3$=$\Oiiib/\hb\ on \muv\ to separate \Oiiib~and \hb. These hypotheses were based on previous works mostly probing brighter galaxies and extrapolations from lower redshift.
Here we report a new statistical sample of \Oiiib\ and \hb\ measurements for galaxies observed with deep medium G395M NIRSpec/JWST spectroscopy behind the strong lensing cluster Abell S1063, which robustly constrains the dependence of R3 and metallicity on \muv\ down to very faint objects at $z \sim 6-7$. This allows us to better quantify the impact of the faintest galaxies on cosmic re-ionisation.

\begin{figure*}[ht]
    \centering
    \includegraphics[width=1\linewidth]{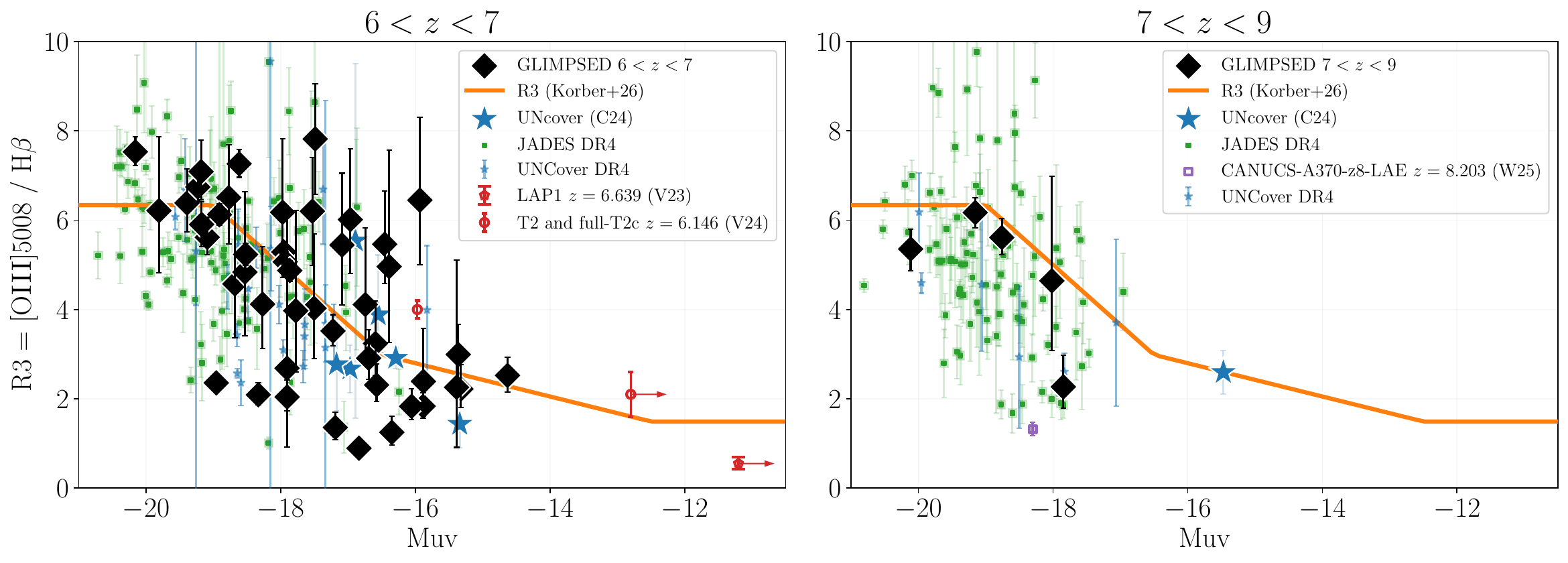}
    \caption{Evolution of R3 ratio as function of \muv\ for the GLIMPSE-D galaxies, as well as the public JADES DR4 \citep{scholtz_jades_2025}, UNCOVER DR4 \citep{price_uncover_2025}, and (C24: \citet{chemerynska_extreme_2024}). We further included the few very metal poor galaxies detected between redshift $6<z<9$ (V23,V24: \citet{vanzella_extremely_2023, vanzella_extreme_2024}; W25: \citet{willott_search_2025}). We also show the evolution as assumed in \cite{korber_glimpse_2026}. Left: $6<z<7$ redshift range, which includes many new objects from GLIMPSE-D at the very faint end. Right: $7<z<9$ redshift range corresponding to the ranges with \oiiihb~in the F444W NIRCam filter.}
    \label{fig:r3_muv}
\end{figure*}

\section{Observations and data}
\label{sec:data}

The GLIMPSE-D program \citep[DD-9223:][]{fujimoto_glimpsed_2025} observed the strong-lensing cluster Abell S1063 with the JWST/NIRSpec MSA instrument using the G395M grating to obtain rest-optical emission-line spectra of some of the intrinsically faintest galaxies at $z \ga 6$ \citep[GLIMPSE,][]{atek_jwst_2025}.
Most of the galaxies were observed for $\sim 9-11$ h, but a subset was observed multiple times, with total integration times up to $\sim 30$ h. The reduction followed \cite{fujimoto_glimpsed_2025} and resulted in 61 galaxies with robust redshifts at $6<z<9$. 
For objects with multiple spectra, we stacked the data by averaging the flux in each wavelength bin and propagating the associated uncertainties.
We visually examined the 1D and 2D spectra, identifying spectra with missing information at the location of \oiii~and \hb.
Seven galaxies were discarded, leaving a final robust sample of 54 galaxies.
The line flux of \oiii\ and \hb\ were measured using three Gaussian profiles, fitted with a standard Markov chain Monte Carlo procedure, as detailed in Appendix\ \ref{app:line_fitting}. 
\muv\ was derived from the GLIMPSE photometric \citep{kokorev_glimpse_2025} catalogue and our spectroscopic redshifts.
All the measured properties are tabulated in Table\ \ref{tab:glimpsed_galaxies}.

To expand the parameter space probed in R3 and \muv\ we also used public data taken from two main surveys: the very deep and large but unlensed field JADES DR4 \citep{scholtz_jades_2025}; and the deep, lensed, but low-volume survey UNCover \citep{bezanson_jwst_2024, suess_medium_2024, furtak_uncovering_2023, price_uncover_2025}. Since some of the sources in JADES were observed in multiple gratings, we measured their average flux and propagated the uncertainties. 
For UNCover we used the values of \citet{chemerynska_extreme_2024} for the faintest galaxies, and we filtered out four clear little red dots around R3$\sim0.1$.

\section{Results and discussions}

\subsection{\Oiiib\ emission and metallicity estimates of the faintest galaxies}\label{sec:o3_Z}

Figure \ref{fig:r3_muv} presents our R3 measurements as a function of \muv\ for two redshift bins. While intended to explore potential evolution across redshift, the sample lacks observations of faint ($\muv \geq -17$) galaxies at $z>7$. Given this limitation, we assumed no evolution over $z=6-9$ in the following analysis. All further R3 (\muv) relationships are based on the full $z=6-9$ dataset.
GLIMPSE-D significantly expands the sample of galaxies with R3 measurements at magnitude fainter than $\muv \gtrsim -16$ at $6<z<7$.
Our measurements show a decline of R3 in fainter galaxies.
The decline in R3 is significant (bootstrapped spearman with N=1000: stat=$-0.42 \pm 0.03$, pval$<0.05$).
This behaviour matches the proposed relationship from \cite{korber_glimpse_2026}, which was based on sparse data.  
At $z \sim 7-9$, the current data are poorly sampled below $-18$ but seem to show a hint of a decline at $z \sim 6-7$ down to $\muv < -17$. In the fainter regime, however, the limited sample of galaxies forces us to use the same dependence as for $z=6-7$.

\begin{figure}[t]
    \centering
    \includegraphics[width=\linewidth]{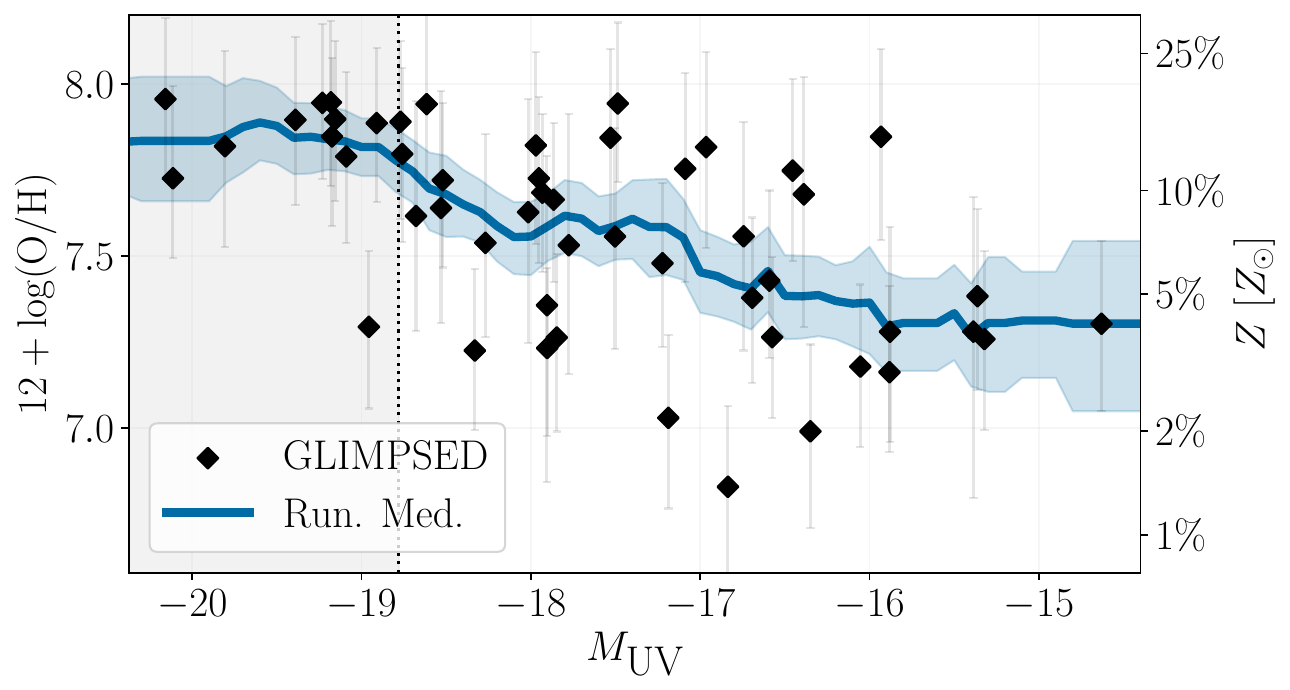}
    \caption{
    Oxygen abundance of selected GLIMPSE-D galaxies using the R3 calibration of \cite{nakajima_empress_2022}. Measurements and uncertainties for individual sources from R3 and the calibration are shown in black, and the oxygen abundance of the R3 running median is shown in blue (16-50-84 percentiles). Metallicities were bootstrapped from $N=1000$ realisations.
    }
    \label{fig:Z}
\end{figure}

The \Oiiib-to-\hb\ ratio (R3) is a well-established metallicity tracer in star-forming galaxies \citep[e.g.][]{sanders_aurora_2025}. For $\oh \la 8.0$, R3 decreases monotonically with metallicity, as confirmed by photoionisation models \citep[e.g.][]{nakajima_empress_2024}.
Qualitatively, the observed decrease of R3 with \muv\ is therefore expected from known mass-metallicity relationships, if \muv\ also traces stellar mass to first order \citep[e.g.][]{song_evolution_2016, rojas-ruiz_borg-jwst_2025}. 
To translate R3 to metallicity, we used the calibrated \Z~measurements from \citet{nakajima_empress_2022} (all samples calibration), assuming that the GLIMPSE-D objects are on the low-metallicity branch of the R3-O/H relation (i.e.~at $\oh \la 8.0$).  
Figure~\ref{fig:Z} shows the resulting metallicities as a function of \muv, spanning oxygen abundances of $\oh \sim 6.9-7.9$ ($\sim 2-20$~\% solar, adopting $\oh_{\rm solar}=8.69$ from \cite{asplund_chemical_2009}).
Objects around $\oh\sim7.9$ have $\Rt\gtrsim6$ at or above the maximum of the R3 metallicity calibration from \citet{nakajima_empress_2022}. 
The lowest abundances are close to the observed metallicity floor at $\oh \sim7.0$ found so far, both in low-$z$ and high-$z$ observations \citep[e.g.][]{sanders_aurora_2025, nakajima_empress_2022}. 
The UV-faint galaxies with extremely low metallicities remain challenging to observe due to their faint nebular features, as shown by recent observations at $z \sim 6$ and other searches for extremely metal-poor galaxies \citep[e.g.][]{morishita_pristine_2025, vanzella_pristine_2026, vanzella_extreme_2024, vanzella_extremely_2023, breneman_leonessa_2025}.
An in-depth analysis of the properties of the faintest ($\muv > -17$) GLIMPSE-D galaxies is presented in \citet{asada_glimpse-ddt_2026}.

\subsection{Impact of the faintest galaxies on re-ionisation}

\begin{figure}[t]
    \centering
    \includegraphics[width=\linewidth]{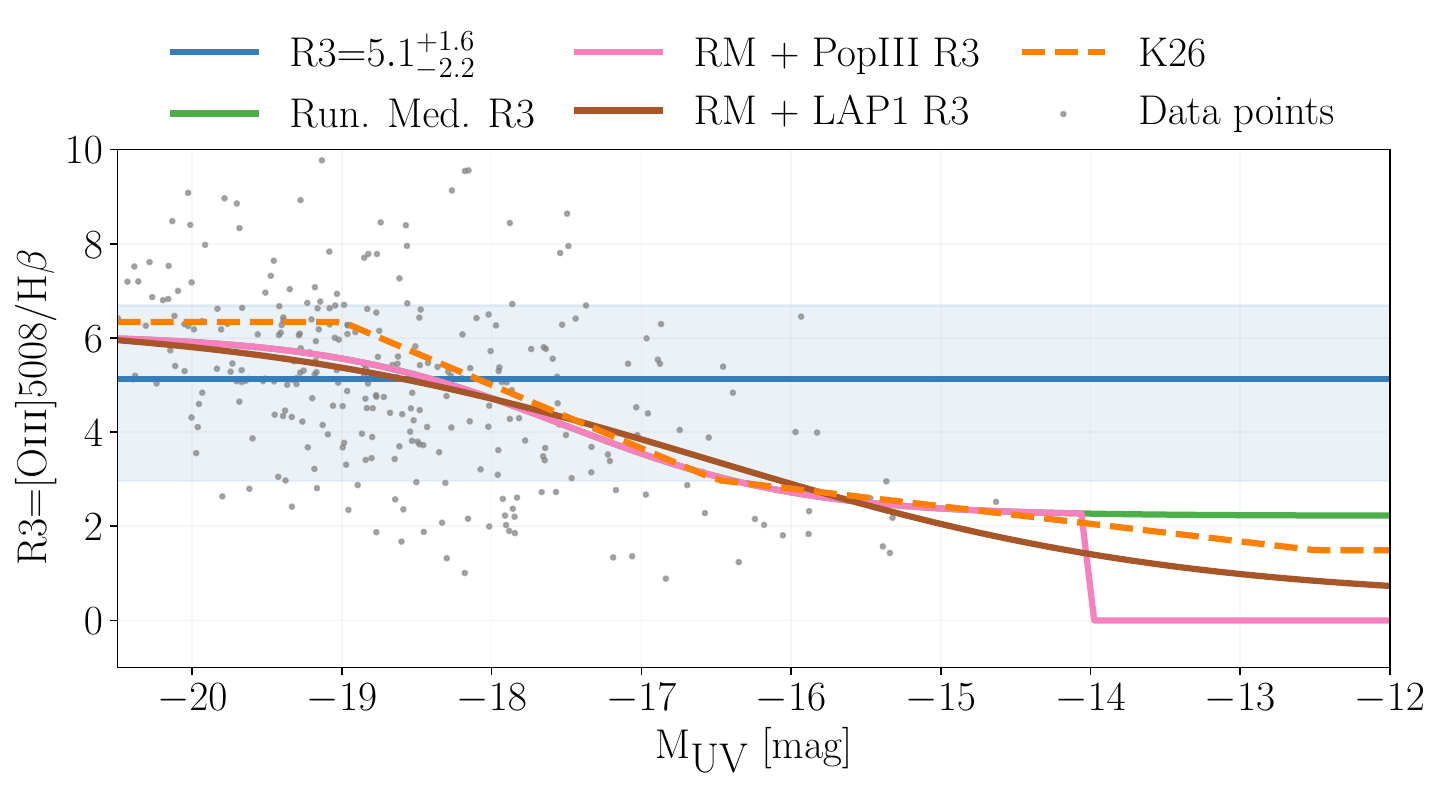}
    \caption{Prescription models of R3(Muv) with the UNCOVER, JADES, and GLIMPSE-D datapoints in the background. Note that the blue area corresponds to the 16th--84th percentiles of the constant model R3.}
    \label{fig:r3muv_models}
\end{figure}

To describe the dependence of the R3 ratio on \muv\ we propose four simple prescription models based on the data available at $6<z<9$. 
Since the sample at $7<z<9$ is small, we assumed that R3(\muv) does not evolve over $6 < z < 9$. These prescriptions are as follows.
{\em i)} A median value of $\Rthree=5.1_{-2.2}^{+1.6}$, independently of \muv. It exhibits a fairly unrealistic view, as the mass-metallicity relationship predicts a decrease of the oxygen abundance for fainter (i.e. lower mass) galaxies \citep[e.g.][]{chemerynska_extreme_2024}, also supported by observations \citep[e.g.][]{vanzella_extremely_2023}, but it provides a test of a homogeneous population in oxygen abundance. 
{\em ii)} A sigmoid fit of the running median. This model follows the observational data, with the assumption of a relatively high faint-end metallicity floor ($\sim 4\%$ solar) due to incompleteness at $\muv \gtrsim -16$.
{\em iii)} The same features as prescription ii) but with a sharp cut-off at $\muv > -14$, representing an extreme scenario where the very-faint galaxy population is dominated by metal-poor galaxies.
{\em iv)} The same features as prescription ii) but with a faint-end metallicity floor fixed to the 16th-84th percentile value of the most-metal poor object observed so far during re-ionisation \citep[LAP1][]{vanzella_extremely_2023}. 
These scenarios are displayed in Fig.~\ref{fig:r3muv_models}. 

Using these models, we can now quantify the contribution of galaxies to cosmic re-ionisation following the procedure from \cite{korber_glimpse_2026}. 
In short, we determine the ionising photon-production rate, \nion\, from their \oiiihb\ LF using our prescription models for \Rthreefunc\ to determine the contribution of \hb\ and, hence, the \hb\ LF. Since hydrogen recombination lines are direct probes of the photoionisation rate, the integral over the \hb\ LF is a direct counter of the total ionising photon-production rate, when a constant escape fraction of ionising photons, \fesc, is assumed. From this, the rate of photons available to re-ionise the IGM can be determined.
See Appendix\ \ref{app:nion} for more details.

\begin{figure*}[t]
    \centering
    \includegraphics[width=1\linewidth]{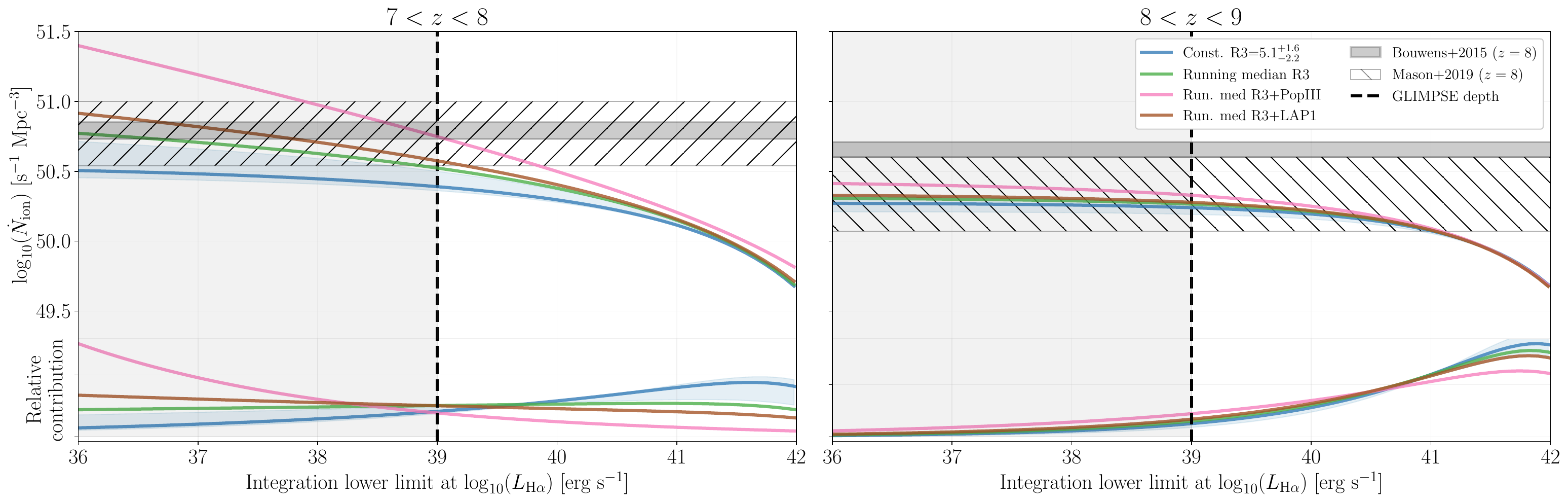}
    \caption{Integrated ionising photon-production rate, \nion\, down to a given $L_{\ha}$, measured from the GLIMPSE \oiiihb~LF with the four R3 models (top panels) and the relative contribution of each luminosity bin to the ionising photon-production rate (bottom panels). We used the average escape fraction measured for GLIMPSE $\fesc=14\%$ \citep{jecmen_glimpse_2026}. The left and right panels differentiate between the LFs from the two redshift bins of \citet{korber_glimpse_2026}. More details in  Appendix\ \ref{app:nion}.}
    \label{fig:nion_models}
\end{figure*}

In Fig.\ \ref{fig:nion_models} we show the cumulative ionising photon rate, \nion\, from the two LFs of \citet{korber_glimpse_2026} ($7<z<8$ and $8<z<9$) using the four prescription models for \Rthreefunc\ and assuming a constant $\fesc=14\%$ \citep{jecmen_glimpse_2026}. In addition, we  show the ionising photon-production rate required to explain the observed ionisation state of the IGM at the considered redshifts, as inferred from Planck and multiple \lya~tracers \citep{mason_model-independent_2019, bouwens_reionization_2015}.

We first focused on $7<z<8$. Due to the constant ionising contribution with model {\em i)}, the \nion\ curve follows the LF and rapidly flattens, effectively undershooting the required budget. {\em ii)} provides a higher contribution of the faintest emitters and reaches the median budget of \citet{mason_model-independent_2019}. Model {\em iii)} strongly differs from the others, predicting a very steep increase of \nion, effectively overshooting the budget, due to the postulated absence of \oiii\ emission in galaxies fainter than $\muv > -14$. Such a scenario requires a drastic decrease of \fesc\ or a sharp decrease of the faint galaxies' number densities to lower the \nion\ budget.

For an integration limit of $L_{\ha}\sim10^{36}$ \ergs, which corresponds to SFR(\ha) $\sim 10^{-5}$ \msunyr, i) undershoots the median budget estimation of \cite{mason_model-independent_2019} by $\sim 0.25$ dex, meaning that a full contribution from galaxies requires a photon-production rate that is a factor of $\sim 1.8$ more ionising. The budget can be reached by involving higher \fesc, but it could also indicate contributions of AGNs.
Models {\em ii)} and {\em iv)}, with an evolving R3, naturally predict more ionising photons from faint galaxies.
Finally, model {\em iv)}, which uses LAP1 \citep{vanzella_extremely_2023} to constrain the R3 of the faintest objects, slightly overshoots the photon budget of \cite{bouwens_reionization_2015}, but it is well within the estimations of \cite{mason_model-independent_2019}. 
These results provide evidence that model {\em iv) \emph{works}}, supporting a scenario where galaxies are the dominant drivers of re-ionisation, with only a minor role for alternative ionising sources. While the current number of detections of objects similar to LAP1 are very limited, this could simply be explained by observational biases as such galaxies require peculiar conditions to be detected. In that scenario, the metallicity floor observed at $\oh\sim7.0$ would be lower than that currently observed \citep[e.g.][]{sanders_aurora_2025, nakajima_empress_2022}.

At redshift $8<z<9$, the very flat faint-end slope of the luminosity function results in all scenarios converging to the same outcome. Each model reaches the median ionising photon budget from \cite{mason_model-independent_2019}, indicating that galaxies are the primary drivers of re-ionisation in this redshift range. However, these estimations undershoot the older estimations from \cite{bouwens_reionization_2015}. Notably, the relative contributions of each luminosity bin indicate that, in these scenarios, the re-ionisation is dominated by galaxies with $L(\ha)\sim10^{41-42}$ \ergs\ and that faint emitters (i.e. $< 10^{40}$ \ergs) contribute only negligibly to re-ionisation.

Finally, we note that our results are derived from a single lensing field, and cosmic variance, particularly for faint galaxies, cannot be quantified with the current limited data (details in Appendix \ref{app:validity}). While consistent with other surveys, these findings are specific to Abell S1063, and broader conclusions will require observations across multiple fields.

\section{Conclusions}
\label{sec:conclusion}
Using deep JWST NIRSpec observations of strongly lensed galaxies behind Abell S1063, we measured the \Rt=\Oiiib/\hb~ratio of 54 very faint ($\sim-20\la\muv<-14$) galaxies.
We found a significant turnover of \Rt\ at $\muv\sim-18$ and a systematic decrease towards fainter UV magnitudes. We interpret the observed decrease as a decrease in the median O/H abundance (metallicity) for fainter galaxies, in agreement with expectations of lower metal enrichment in fainter and less massive galaxies \citep[e.g.][]{sanders_aurora_2025}. 
We found metallicities down to $\sim 2$ \% solar at $\muv\sim-17$, close to the lowest metallicities observed in galaxies.

Using the observational data, we built four prescription models of the evolution of \Rt\ with \muv\ and inferred the contribution of galaxies to the ionising photon production behind Abell S1063 using the GLIMPSE \oiiihb\ LF \citep{korber_glimpse_2026}.
Extreme models ({\em i, iii}) either under- or overshoot the ionising photon budget from \citet{mason_model-independent_2019}, while intermediate models ({\em ii, iv}) align with it, reducing the budget compared to earlier estimates \citep[e.g.][]{munoz_reionization_2024}.
Finally, our results are consistent with a re-ionisation driven by star-forming galaxies in the Abell S1063 region, and we thus argue in favour of the existence of faint and metal-poor galaxies such as LAP1 \citep{vanzella_extremely_2023}. We show that the hypotheses on the decline of the average R3 with fainter \muv\ from \citet{korber_glimpse_2026} hold and are compatible with a lower metallicity floor, as shown by model {\em iv)}. These results suggest that the faintest galaxies detected by GLIMPSE do not contribute significantly to the ionising budget during re-ionisation. 

\section*{Data availability}
Table \ref{tab:glimpsed_galaxies} is available in electronic form at the CDS via \href{https://cdsarc.cds.unistra.fr/viz-bin/cat/J/A+A/710/L15}{cdsarc.cds.unistra.fr/viz-bin/cat/J/A+A/710/L15}.

\bibliographystyle{aa}
\bibliography{references.bib}

\begin{appendix}

\section{Acknowledgements}
\begin{acknowledgements} 
DK thanks Romain Meyer, Andrea Weibel, Lucie Scharré, Danielle Berg and Ryan Sanders for the useful discussions on the project. AA acknowledges support by the Swedish research council Vetenskapsr{\aa}det (VR) project 2021-05559, and VR consolidator grant 2024-02061. This work is based [in part] on observations made with the NASA/ESA/CSA James Webb Space Telescope. The data were obtained from the Mikulski Archive for Space Telescopes at the Space Telescope Science Institute, which is operated by the Association of Universities for Research in Astronomy, Inc., under NASA contract NAS 5-03127 for JWST. 
These observations are associated with programs \#9223 and \#3293.
We acknowledge the support of the Canadian Space Agency (CSA) [25JWGO4A06]. 
SF acknowledges support from the Dunlap Institute, funded through an endowment established by the David Dunlap family and the University of Toronto. 
Softwares: Emcee \citep{foreman-mackey_emcee_2013}, Astropy \citep{collaboration_astropy_2022}, Numpy \citep{harris_array_2020}, Scipy \citep{virtanen_scipy_2020}, Matplotlib \citep{hunter_matplotlib_2007}.
\end{acknowledgements} 

\section{Statistical validation of the trend}\label{app:validity}

Our results are derived from a single strongly lensed field, which correspond to a relatively small volume \citep[c.f. see Fig. 2 in ][]{korber_glimpse_2026}. Therefore, our results are valid behind Abell S1063, but requires extra care for generalisation. 

\subsection{Cosmic variance}

\begin{figure}[t]
    \centering
    \includegraphics[width=\linewidth]{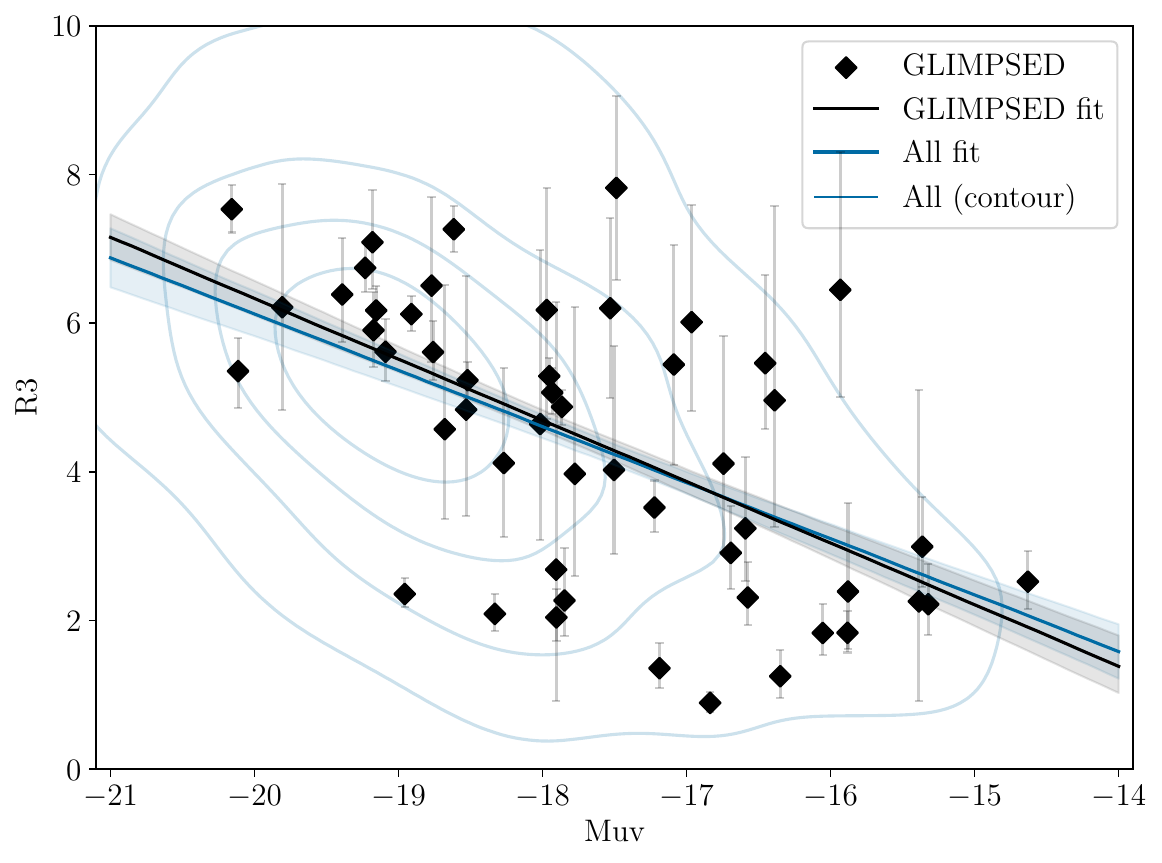}
    \caption{Linear fit of the $\muv<-14$ data from GLIMPSED (black star for the data and black line for the linear fit) and the data from GLIMPSED, UNCOVER and JADES, as reported in this letter (blue contour and blue line for the linear fit).}
    \label{fig:r3_linear_fit}
\end{figure}

The results rely on the observed decreasing trend of the $R3$ ratio for fainter galaxies, though this trend may be influenced by cosmic variance, i.e. field-to-field variation. Estimating cosmic variance for the faintest galaxies is particularly challenging due to the scarcity of comparable data. Nevertheless, in Fig.~\ref{fig:r3_linear_fit}, we show the linear fit of the GLIMPSE-D data (black) alongside data from JADES, UNCOVER, and GLIMPSE-D, which cover distinct fields (GOODS-N/S, Abell~2744, and Abell~S1063). No significant differences are observed between the fits, suggesting that the effect of cosmic variance on the relationship might be limited. Nonetheless, future surveys of deep lensed fields are required to assess on the impact of cosmic variance on this relationship.

Additionally, complementing the Spearman correlation presented in Sect.~\ref{sec:o3_Z}, the linear fits reveal a significant decline of $R3$ with \muv. This aligns with expectations from the mass-metallicity relation \citep[e.g.][]{chemerynska_extreme_2024}, as discussed earlier. We note that the linear fit is a first-order approximation and does not apply to the faintest and brightest objects ($R3 < 0$ and $R3 \geq 10$); thus, the fit is restricted to $M_{\text{UV}} < -14$.

\subsection{Completeness}
The GLIMPSE-D sample is incomplete, so it remains unclear whether we are probing typical galaxies or outliers. While we have assembled a relatively large sample at these magnitudes, we have already demonstrated the impact of different prescription models on the results of \citep{korber_glimpse_2026}. Below, we discuss potential galaxies that may have been missed.

At the bright end ($\muv\leq-20$), we assume all star-forming galaxies would be detected regardless of their $R3$ ratio, and other survey suggest that galaxies with $R3<4$ at these magnitudes are uncommon at such redshifts.

At the very faint end ($\muv\geq-16$), galaxies do not always exhibit detected continuum and may only be observable via emission lines. We therefore divide the analysis into two cases:
\begin{itemize}
    \item High $R3\geq4$ ratios: These could arise from strong \oiii\ emission or weak \hb\ emission. In the former case, galaxies would likely be detected, as \oiii\ is among the strongest lines at high redshift. In the latter, high $R3$ ratios would carry large uncertainties, e.g. as seen in ID=46408. Strong \oiii\ emission was not observed in our faintest galaxies, but this may simply reflect the mass–metallicity relation, so we do not expect a significant bias here.
    \item Low $R3\leq1$ ratios: Faint galaxies with undetected continuum and low $R3$ ratios may be difficult to observe. Currently, only a handful of such galaxies are known, which could bias our results upward. This domain of the parameter space remains unexplored.
\end{itemize}

To address these possible biases, we include prescription models {\em i}) and {\em iii}), which serve as extreme boundaries models.

\section{Fitting emission lines}
\label{app:line_fitting}

\begin{figure*}[t]
    \centering
    \includegraphics[width=\linewidth]{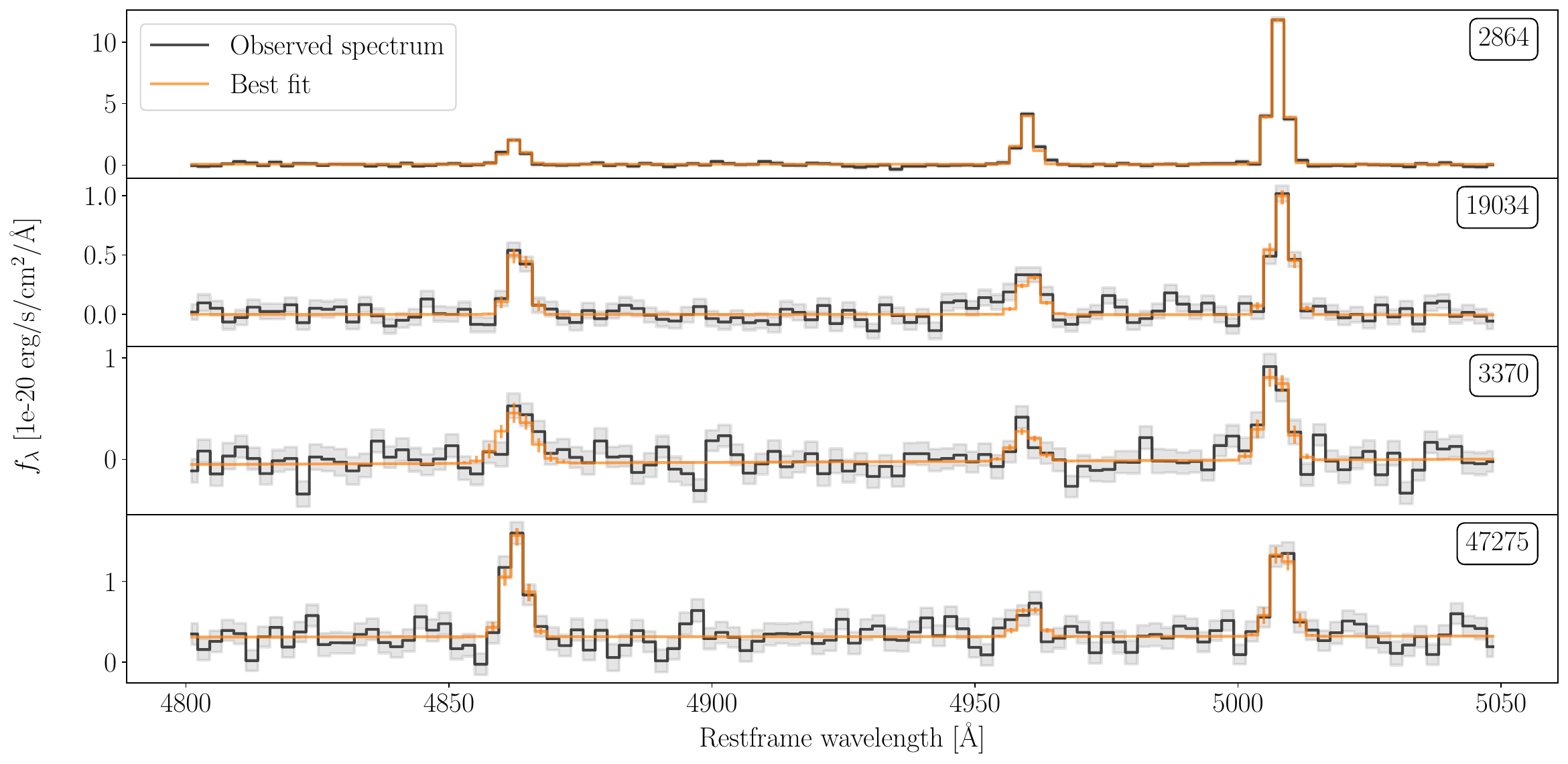}
    \caption{Four galaxies with their observed spectrum and their associated best fit model. The galaxies are ordered by their R3 (from highest to lowest).}
    \label{fig:examples}
\end{figure*}

To measure the line flux of the \oiii\ and \hb\ lines, we first fitted the continuum. For that, we selected small regions around the lines and masked them. We also manually checked each spectrum and removed contaminating features from other objects. Then, we fitted a 1st order polynomial to the continuum, which was then subtracted from the spectrum.
To model the emission lines, we use one Gaussian profile per line. For \oiiihb, this yield nine free parameters, which can be reduced with three physical considerations: The rest-frame wavelengths are fixed by physics and can be reduced to one physical parameter governed by the redshift. Then, the amplitude between the two \oiii\ doublet is fixed to 2.98 \citep{storey_theoretical_2000}, reducing them to one free parameter. Finally, we expect the \oiii\ doublet to come from the same region, and therefore have the same velocity profile, so the standard deviation is common to the two Gaussians.
We end up with five free parameters ($z$, $A_{\hb}$, $A_{\Oiiib}$, $\sigma_{\hb}$ and $\sigma_{\Oiiib}$, where $A$ are the Gaussian amplitudes and $\sigma$ are standard deviations), that we optimise with a Monte Carlo Markov chain (MCMC). Note that the spectra of two galaxies show broader components. We fitted only the narrow component, leaving the detailed analysis for other GLIMPSE papers (see \citet{berg_fleeting_2025} and \citet{korber_direct_2026}).  
Figure\ \ref{fig:examples} shows four galaxies with their associated fits.
The measured line flux are found in Table\ \ref{tab:glimpsed_galaxies}. 

\section{Measuring \nion~from the \oiiihb~LF}
\label{app:nion}
The ionising photon production rate \nion~can be directly obtained from the \ha~LF (or \hb\ LF, accounting for dust attenuation and a fixed \ha-to-\hb\ ratio) $\phi$ following

\begin{align}
    \dot{N}_\text{ion} &= \int_{L_\text{min}}^\infty \fesc Q_\text{ion}(L)\phi(L) dL\\
    &= \int_{L_\text{min}}^\infty \fesc\frac{L}{c_\alpha (1-\fesc)}\phi(L) dL.
    \label{eq:emissivity}
\end{align}

\noindent For this, we assumed a constant $\fesc=14\%$, following the measurements of \citet{jecmen_glimpse_2026} for the GLIMPSE photometric sample. Their measurements span galaxies down to $\muv\gtrsim-12.5$ using the UV continuum slopes.
We also adopted the ratio between the line emissivity and the total recombination rate $c_\alpha = 1.37\times10^{-12}$ erg for an electron temperature of $10^4$K \citep{schaerer_transition_2003}. The ionising photon production rate $Q_{\rm ion}(L)$ is related to luminosity with $L=Q_{\rm ion}c_\alpha(1-\fesc)$.
The relative contribution of each luminosity bin in Fig.\ \ref{fig:nion_models} is measured by calculating the difference between all the \nion\ data point and normalising it by the \nion\ at $L_{\ha}=10^{36}$\ s$^{-1}$Mpc$^{-3}$.

\section{Table of emission lines}

We report in Table \ref{tab:glimpsed_galaxies} the measurements of the galaxies considered in this Letter.

\begin{table*}[ht]
    \centering
\caption{Catalogue and properties of the sources}
    \label{tab:glimpsed_galaxies}
\begin{tabular}{c|c|c|c|c|c|c|c|c|c}
       \toprule
        ID & RA & DEC & $z_{\rm spec}$ & \muv & $\mu$ & \Oiiib  & \hb & R3 & 12+log(O/H) \\
         & deg & deg & & AB mag & & $10^{-19}$ egs & $10^{-19}$ egs &  \\
        (1) & (2) & (3) & (4) & (5) & (6) & (7) & (8) & (9) & (10) \\ 
        \midrule
    2401 & 342.22476 & -44.56736 & 7.9235 & -20.11 & $1.37 \pm 0.02$ & $34.28_{-0.97}^{+0.77}$ & $6.29_{-0.60}^{+0.47}$ & $5.35_{-0.49}^{+0.44}$ & $7.73_{-0.23}^{+0.27}$\\
2864 & 342.23004 & -44.56598 & 6.4890 & -17.93 & $1.28 \pm 0.01$ & $26.21_{-0.40}^{+0.40}$ & $5.17_{-0.30}^{+0.30}$ & $5.07_{-0.29}^{+0.32}$ & $7.68_{-0.23}^{+0.23}$\\
3370 & 342.22861 & -44.56467 & 6.1738 & -17.19 & $1.29 \pm 0.02$ & $2.91_{-0.33}^{+0.35}$ & $2.14_{-0.38}^{+0.43}$ & $1.36_{-0.27}^{+0.34}$ & $7.03_{-0.26}^{+0.24}$\\
4164 & 342.21524 & -44.56320 & 6.1757 & -18.53 & $1.39 \pm 0.03$ & $5.82_{-0.34}^{+0.30}$ & $1.18_{-0.34}^{+0.46}$ & $4.84_{-1.42}^{+1.80}$ & $7.64_{-0.33}^{+0.34}$\\
5381 & 342.26215 & -44.56067 & 7.5492 & -19.15 & $1.22 \pm 0.01$ & $61.29_{-0.78}^{+0.73}$ & $9.93_{-0.48}^{+0.57}$ & $6.17_{-0.35}^{+0.33}$ & $7.90_{-0.24}^{+0.23}$\\
5536 & 342.25626 & -44.56019 & 6.2228 & nan & $1.23 \pm 0.01$ & $5.02_{-0.22}^{+0.22}$ & $1.11_{-0.20}^{+0.20}$ & $4.46_{-0.73}^{+0.95}$ & $7.60_{-0.26}^{+0.24}$\\
6170 & 342.25748 & -44.55910 & 6.4367 & -18.62 & $1.23 \pm 0.01$ & $31.23_{-0.37}^{+0.38}$ & $4.30_{-0.17}^{+0.18}$ & $7.26_{-0.30}^{+0.32}$ & $7.94_{-0.23}^{+0.27}$\\
6297 & 342.21240 & -44.55897 & 7.1221 & nan & $1.44 \pm 0.03$ & $42.49_{-1.55}^{+1.54}$ & $12.98_{-1.48}^{+1.52}$ & $3.28_{-0.36}^{+0.44}$ & $7.41_{-0.22}^{+0.25}$\\
6358 & 342.26428 & -44.55873 & 6.2252 & -18.96 & $1.22 \pm 0.01$ & $9.62_{-0.32}^{+0.31}$ & $4.07_{-0.32}^{+0.31}$ & $2.36_{-0.18}^{+0.21}$ & $7.29_{-0.24}^{+0.22}$\\
6408 & 342.26740 & -44.55862 & 6.4388 & -17.53 & $1.21 \pm 0.01$ & $11.09_{-0.37}^{+0.28}$ & $1.76_{-0.30}^{+0.36}$ & $6.20_{-1.21}^{+1.21}$ & $7.84_{-0.27}^{+0.26}$\\
6587 & 342.26785 & -44.55857 & 6.4380 & -19.39 & $1.21 \pm 0.01$ & $30.61_{-0.56}^{+0.56}$ & $4.78_{-0.50}^{+0.51}$ & $6.39_{-0.64}^{+0.76}$ & $7.90_{-0.25}^{+0.24}$\\
7468 & 342.25397 & -44.55680 & 6.1124 & -19.23 & $1.25 \pm 0.02$ & $45.28_{-0.67}^{+0.68}$ & $6.71_{-0.31}^{+0.32}$ & $6.74_{-0.32}^{+0.35}$ & $7.94_{-0.22}^{+0.23}$\\
7685 & 342.24686 & -44.55608 & 6.4175 & -16.59 & $1.28 \pm 0.02$ & $3.41_{-0.26}^{+0.25}$ & $1.05_{-0.25}^{+0.29}$ & $3.24_{-0.71}^{+0.96}$ & $7.43_{-0.22}^{+0.26}$\\
8139 & 342.26129 & -44.55530 & 6.2238 & -16.45 & $1.24 \pm 0.01$ & $7.48_{-0.29}^{+0.29}$ & $1.37_{-0.24}^{+0.24}$ & $5.46_{-0.88}^{+1.19}$ & $7.75_{-0.26}^{+0.27}$\\
8172 & 342.24261 & -44.55537 & 6.2199 & -17.49 & $1.29 \pm 0.02$ & $31.88_{-0.81}^{+0.78}$ & $3.97_{-0.66}^{+0.79}$ & $7.82_{-1.24}^{+1.24}$ & $7.94_{-0.23}^{+0.24}$\\
8192 & 342.24255 & -44.55531 & 6.2199 & -18.27 & $1.29 \pm 0.02$ & $16.62_{-0.56}^{+0.52}$ & $3.98_{-0.91}^{+1.18}$ & $4.12_{-0.99}^{+1.28}$ & $7.54_{-0.27}^{+0.32}$\\
8947 & 342.26410 & -44.55429 & 6.2208 & -18.77 & $1.23 \pm 0.01$ & $17.10_{-0.59}^{+0.54}$ & $2.62_{-0.41}^{+0.48}$ & $6.50_{-1.03}^{+1.19}$ & $7.89_{-0.28}^{+0.23}$\\
8981 & 342.18991 & -44.55407 & 6.5588 & -19.09 & $1.61 \pm 0.03$ & $37.45_{-0.79}^{+0.80}$ & $6.67_{-0.47}^{+0.49}$ & $5.62_{-0.39}^{+0.44}$ & $7.79_{-0.25}^{+0.25}$\\
9081 & 342.25659 & -44.55378 & 6.2222 & -16.97 & $1.26 \pm 0.01$ & $6.90_{-0.29}^{+0.29}$ & $1.14_{-0.25}^{+0.29}$ & $6.01_{-1.20}^{+1.58}$ & $7.82_{-0.29}^{+0.28}$\\
9214 & 342.25241 & -44.55355 & 6.5491 & -17.22 & $1.27 \pm 0.02$ & $8.62_{-0.23}^{+0.21}$ & $2.43_{-0.23}^{+0.23}$ & $3.52_{-0.33}^{+0.36}$ & $7.48_{-0.24}^{+0.23}$\\
9751 & 342.19452 & -44.55263 & 6.5110 & -16.39 & $1.77 \pm 0.05$ & $2.40_{-0.22}^{+0.23}$ & $0.40_{-0.23}^{+0.27}$ & $4.96_{-1.70}^{+2.61}$ & $7.68_{-0.39}^{+0.34}$\\
10182 & 342.24286 & -44.55198 & 6.5175 & -17.90 & $1.31 \pm 0.02$ & $5.10_{-0.35}^{+0.33}$ & $2.49_{-0.37}^{+0.39}$ & $2.04_{-0.32}^{+0.38}$ & $7.23_{-0.25}^{+0.23}$\\
10357 & 342.24915 & -44.55183 & 6.0600 & -18.91 & $1.29 \pm 0.02$ & $84.80_{-1.14}^{+1.14}$ & $13.86_{-0.49}^{+0.51}$ & $6.12_{-0.23}^{+0.24}$ & $7.89_{-0.23}^{+0.22}$\\
11482 & 342.26190 & -44.54998 & 6.5060 & -16.74 & $1.26 \pm 0.01$ & $5.35_{-0.31}^{+0.34}$ & $1.29_{-0.39}^{+0.51}$ & $4.11_{-1.19}^{+1.72}$ & $7.56_{-0.33}^{+0.33}$\\
11525 & 342.24811 & -44.54990 & 6.5049 & -18.33 & $1.31 \pm 0.02$ & $5.49_{-0.28}^{+0.27}$ & $2.63_{-0.29}^{+0.29}$ & $2.09_{-0.23}^{+0.27}$ & $7.23_{-0.23}^{+0.24}$\\
11789 & 342.25421 & -44.54950 & 6.5211 & -16.69 & $1.28 \pm 0.02$ & $7.65_{-0.47}^{+0.44}$ & $2.61_{-0.49}^{+0.48}$ & $2.91_{-0.48}^{+0.63}$ & $7.38_{-0.25}^{+0.23}$\\
12248 & 342.23199 & -44.54891 & 6.1081 & -17.87 & $1.40 \pm 0.02$ & $34.89_{-0.80}^{+0.71}$ & $7.16_{-0.28}^{+0.32}$ & $4.87_{-0.25}^{+0.22}$ & $7.66_{-0.24}^{+0.22}$\\
12801 & 342.24939 & -44.54817 & 7.5686 & -18.76 & $1.31 \pm 0.02$ & $28.49_{-0.51}^{+0.52}$ & $5.08_{-0.35}^{+0.36}$ & $5.61_{-0.38}^{+0.42}$ & $7.80_{-0.26}^{+0.25}$\\
13186 & 342.17361 & -44.54759 & 6.1571 & -16.35 & $2.90 \pm 0.09$ & $1.04_{-0.14}^{+0.15}$ & $0.83_{-0.17}^{+0.20}$ & $1.25_{-0.29}^{+0.36}$ & $6.99_{-0.28}^{+0.25}$\\
13882 & 342.23651 & -44.54689 & 6.1074 & -19.17 & $1.39 \pm 0.02$ & $46.52_{-1.15}^{+1.03}$ & $7.81_{-0.67}^{+0.61}$ & $5.90_{-0.50}^{+0.51}$ & $7.85_{-0.26}^{+0.23}$\\
13941 & 342.23645 & -44.54690 & 6.1075 & -19.18 & $1.39 \pm 0.03$ & $18.63_{-0.44}^{+0.44}$ & $2.63_{-0.23}^{+0.25}$ & $7.09_{-0.63}^{+0.71}$ & $7.95_{-0.24}^{+0.24}$\\
14311 & 342.17770 & -44.54694 & 6.0027 & -17.97 & $3.54 \pm 0.12$ & $6.13_{-0.37}^{+0.33}$ & $0.94_{-0.25}^{+0.25}$ & $6.18_{-1.46}^{+1.65}$ & $7.82_{-0.29}^{+0.27}$\\
14499 & 342.25143 & -44.54671 & 6.1063 & -20.16 & $1.31 \pm 0.02$ & $106.60_{-1.47}^{+1.49}$ & $14.16_{-0.56}^{+0.58}$ & $7.53_{-0.31}^{+0.33}$ & $7.96_{-0.23}^{+0.24}$\\
17070 & 342.17825 & -44.54387 & 6.0296 & -15.32 & $5.80 \pm 0.28$ & $0.61_{-0.06}^{+0.06}$ & $0.28_{-0.05}^{+0.06}$ & $2.22_{-0.41}^{+0.55}$ & $7.26_{-0.26}^{+0.25}$\\
19034 & 342.21237 & -44.52876 & 6.2025 & -15.88 & $2.93 \pm 0.15$ & $1.25_{-0.11}^{+0.10}$ & $0.67_{-0.09}^{+0.09}$ & $1.83_{-0.26}^{+0.30}$ & $7.16_{-0.23}^{+0.25}$\\
26653 & 342.18408 & -44.53164 & 6.1044 & -19.81 & $1.96 \pm 0.36$ & $130.85_{-20.15}^{+28.92}$ & $20.76_{-3.19}^{+4.59}$ & $6.21_{-1.39}^{+1.66}$ & $7.82_{-0.29}^{+0.28}$\\
37914 & 342.18845 & -44.53619 & 6.1096 & -15.36 & $16.29 \pm 1.62$ & $1.03_{-0.10}^{+0.12}$ & $0.35_{-0.06}^{+0.06}$ & $2.99_{-0.54}^{+0.67}$ & $7.38_{-0.27}^{+0.25}$\\
38657 & 342.24072 & -44.53654 & 6.2117 & -17.78 & $1.47 \pm 0.03$ & $5.26_{-0.45}^{+0.44}$ & $1.24_{-0.51}^{+0.64}$ & $3.97_{-1.37}^{+2.24}$ & $7.53_{-0.38}^{+0.38}$\\
38807 & 342.23361 & -44.53667 & 7.7668 & -18.02 & $1.59 \pm 0.03$ & $8.40_{-0.52}^{+0.51}$ & $1.71_{-0.68}^{+0.79}$ & $4.64_{-1.56}^{+2.34}$ & $7.63_{-0.38}^{+0.33}$\\
41948 & 342.19086 & -44.53749 & 6.1044 & nan & $6.22 \pm 0.42$ & $145.79_{-9.40}^{+10.54}$ & $22.47_{-1.44}^{+1.65}$ & $6.49_{-0.61}^{+0.66}$ & $7.87_{-0.22}^{+0.25}$\\
42531 & 342.22708 & -44.53763 & 6.2075 & -17.91 & $1.67 \pm 0.04$ & $29.08_{-24.30}^{+2.67}$ & $0.36_{-0.25}^{+0.50}$ & $2.68_{-1.76}^{+3.60}$ & $7.36_{-0.51}^{+0.43}$\\
42812 & 342.24670 & -44.53885 & 6.5932 & -18.68 & $1.39 \pm 0.03$ & $8.09_{-0.48}^{+0.49}$ & $1.74_{-0.55}^{+0.63}$ & $4.57_{-1.20}^{+1.94}$ & $7.62_{-0.33}^{+0.33}$\\
\bottomrule
\multicolumn{10}{l}{\textit{Table continued on next page...}}\\
\end{tabular}
\end{table*}

\begin{table*}[ht]
    \centering
    \begin{tabular}{c|c|c|c|c|c|c|c|c|c}
    \toprule
        ID & RA & DEC & $z_{\rm spec}$ & \muv & $\mu$ & \Oiiib  & \hb & R3 & 12+log(O/H) \\
         & deg & deg & & AB mag & & $10^{-19}$ egs & $10^{-19}$ egs &  \\
        (1) & (2) & (3) & (4) & (5) & (6) & (7) & (8) & (9) & (10) \\ 
        \midrule
        
43280 & 342.24637 & -44.53909 & 6.2242 & -17.50 & $1.39 \pm 0.02$ & $4.48_{-0.35}^{+0.32}$ & $1.09_{-0.33}^{+0.41}$ & $4.03_{-1.13}^{+1.67}$ & $7.56_{-0.33}^{+0.31}$\\
43555 & 342.26297 & -44.53924 & 6.2064 & -17.09 & $1.30 \pm 0.02$ & $6.86_{-0.38}^{+0.34}$ & $1.25_{-0.30}^{+0.37}$ & $5.44_{-1.35}^{+1.61}$ & $7.75_{-0.33}^{+0.28}$\\
45084 & 342.25854 & -44.54039 & 6.4940 & -16.58 & $1.31 \pm 0.02$ & $3.34_{-0.24}^{+0.24}$ & $1.45_{-0.25}^{+0.25}$ & $2.31_{-0.37}^{+0.48}$ & $7.26_{-0.24}^{+0.23}$\\
45706 & 342.23682 & -44.54075 & 6.4150 & -16.05 & $1.46 \pm 0.03$ & $3.75_{-0.44}^{+0.34}$ & $2.03_{-0.40}^{+0.33}$ & $1.83_{-0.30}^{+0.39}$ & $7.18_{-0.23}^{+0.24}$\\
45883 & 342.24643 & -44.54088 & 6.2220 & -17.95 & $1.38 \pm 0.02$ & $22.14_{-0.42}^{+0.40}$ & $4.19_{-0.17}^{+0.19}$ & $5.29_{-0.25}^{+0.24}$ & $7.73_{-0.25}^{+0.24}$\\
46408 & 342.18414 & -44.54120 & 6.5361 & -15.93 & $6.29 \pm 0.39$ & $0.84_{-0.06}^{+0.07}$ & $0.12_{-0.04}^{+0.04}$ & $6.45_{-1.45}^{+1.86}$ & $7.85_{-0.30}^{+0.25}$\\
46431 & 342.23651 & -44.54118 & 6.5221 & -15.88 & $1.45 \pm 0.02$ & $2.35_{-0.26}^{+0.25}$ & $0.95_{-0.31}^{+0.38}$ & $2.39_{-0.77}^{+1.18}$ & $7.28_{-0.32}^{+0.30}$\\
46938 & 342.23605 & -44.54156 & 6.8526 & -18.52 & $1.46 \pm 0.03$ & $36.01_{-0.85}^{+0.79}$ & $6.87_{-0.31}^{+0.28}$ & $5.23_{-0.24}^{+0.25}$ & $7.72_{-0.25}^{+0.22}$\\
47275 & 342.24191 & -44.54156 & 6.4427 & -16.84 & $1.40 \pm 0.02$ & $2.99_{-0.33}^{+0.32}$ & $3.36_{-0.34}^{+0.33}$ & $0.89_{-0.12}^{+0.14}$ & $6.83_{-0.27}^{+0.23}$\\
47757 & 342.17593 & -44.54240 & 6.0014 & -14.63 & $14.67 \pm 1.40$ & $0.80_{-0.07}^{+0.09}$ & $0.32_{-0.03}^{+0.04}$ & $2.52_{-0.37}^{+0.41}$ & $7.30_{-0.25}^{+0.24}$\\
54637 & 342.20590 & -44.51908 & 6.0647 & -15.39 & $16.12 \pm 8.09$ & $9.46_{-6.87}^{+13.26}$ & $10.23_{-7.24}^{+14.16}$ & $2.26_{-1.35}^{+2.85}$ & $7.28_{-0.48}^{+0.39}$\\
56616 & 342.22980 & -44.54204 & 8.6354 & -17.85 & $1.52 \pm 0.04$ & $0.23_{-0.08}^{+0.21}$ & $0.09_{-0.06}^{+0.11}$ & $2.27_{-0.47}^{+0.71}$ & $7.26_{-0.27}^{+0.24}$\\
\bottomrule
\end{tabular}
    \tablefoot{(1) GLIMPSE identification number (2 \& 3) Right Ascention and DEClination in J2000 (4) spectroscopic redshift from emission line fit (5) intrinsic UV magnitude. Note that sources with missing \muv~have a poor photometric constraints and were not included in the results. (6) Lensing magnification (7 \& 8) intrinsic line flux of \Oiiib~and \hb\ (egs = erg/s/cm$^2$) (9) $\Rt = \Oiiib/\hb$ (10) Oxygen abundance from \citep{nakajima_empress_2022}. Line fluxes and \muv~were corrected for lensing magnification. To retrieve the observed quantities, multiple line flux by $\mu$ and subtract $2.5\log_{10}(\mu)$ to \muv.}
\end{table*}

\end{appendix}

\end{document}